\documentclass[preprint,showpacs,preprintnumbers,amsmath,amssymb]{revtex4}
\usepackage{graphicx}
\usepackage{dcolumn}
\usepackage{bm}

\begin{document}
\title{Twist-stretch profiles of DNA chains}

\author{Marco Zoli}

\affiliation{School of Science and Technology \\  University of Camerino, I-62032 Camerino, Italy \\ marco.zoli@unicam.it}

\date{\today}

\begin{abstract}
Helical molecules change their twist number under the effect of a mechanical load. We study the twist-stretch relation for a set of short DNA molecules modeled by a mesoscopic Hamiltonian. Finite temperature path integral techniques are applied to generate a large ensemble of possible configurations for the base pairs of the sequence. The model also accounts for the bending and twisting fluctuations between adjacent base pairs along the molecules stack. Simulating a broad range of twisting conformation, we compute the helix structural parameters by averaging over the ensemble of base pairs configurations. The method selects, for any applied force, the average twist angle which minimizes the molecule's free energy. It is found that the chains generally over-twist under an applied stretching and the over-twisting is physically associated to the contraction of the average helix diameter, i.e. to the damping of the base pair fluctuations. Instead, assuming that the maximum amplitude of the bending fluctuations may decrease against the external load, the DNA molecule first over-twists for weak applied forces and then untwists above a characteristic force value. Our results are discussed in relation to available experimental information albeit for kilo-base long molecules.
\end{abstract}


\maketitle

\section*{1. Introduction}

While DNA packaging in bacteria and eukaryotic chromosomes rely on the mechanical properties of the double helix, the sequence specific chain deformability is essential as much to the formation of transient fluctuational bubbles involved in cell division and transcription as to DNA recognition by proteins which twist, bend and stretch the helix upon binding \cite{travers,kalos11,cherstvy}. Proteins, typically interacting with short base pair sequences, say 10 - 20 bps, confer a manifold of functions to the bare DNA chains \cite{marko15}. A remarkable example is provided by recombinases, such as RecA in \textit{E. coli}, which promote homologous pairing between e.g., a partially single stranded DNA and a target double stranded molecule \cite{radding} thus producing displacement and exchange of strands, a vital process in the recombinational DNA repair \cite{cox}. The mechanisms by which the RecA nucleoprotein filament efficiently samples the target dsDNA and pairs homologous sequences are largely unknown: some studies \cite{dekker} suggest that the RecA - dsDNA binding is favored by a double helix unwinding with opening of thermal bubbles whereas others \cite{forget} point out that the rate of homologous pairing is higher when dsDNA maintains a random coil conformation with short end-to-end distance as this enhances the local DNA concentration and allows a rapid homologous search. It is however likely that the specific twist configuration of the helix affects both the size and the binding properties of the recombinase - dsDNA complexes \cite{stasiak} while, in turn, the latter restructure and deform the double helix in several ways \cite{olson98,marko16}.

Considerable knowledge of the DNA properties has been gained over the last twenty five years as a wide number of methods to micro-manipulate single molecules and sample their mechanical response to external forces has become available \cite{chu,busta92,cluzel,block97,das14}.  First measurements on $\lambda$-phage molecules in solution \cite{busta92}  have shown that, for low and intermediate forces, DNA  displays entropic elastic behavior dominated by thermal bending fluctuations which constantly deform base pairs and molecule backbone so that the helix assumes different random walk configurations. Accordingly, inextensible worm-like-chain models with persistence length of $\sim 50$ nm fit well the force-extension data up to applied forces of $\sim 10$ pN \cite{busta94,odi}. Above this typical force, the  polymer end-to-end distance stretches linearly and takes values even larger than the B-form contour length. At $\sim 65$ pN, the force-extension plot shows a peculiar plateau signaling that the molecule undergoes a structural transition and over-stretches to $\sim 1.7$ times its B-form contour length \cite{busta96,marko97,strick98,nelson03}, the precise values depending on the solution condition \cite{rouzina01}. Note that the RecA - dsDNA complex is stretched by a factor 1.5 with respect to the naked DNA length \cite{cassuto}. Importantly, assuming an experimental setup which anchors both ends of the molecule thus preventing it to twist, the over-stretching transition shifts at $\sim 100$ pN \cite{marko99}. Moreover, regardless of the attachment geometry,  optical tweezers methods combined with single molecule fluorescence imaging have visualized that over-stretched dsDNA gradually converts into single stranded DNA \cite{mameren} thus suggesting a force induced melting transition. 

Ten years ago  Gore et al. \cite{busta06} showed that DNA molecules in the kilo-base range over-twist under small to moderate applied tensions (up to  of $\sim 30$ pN) and eventually untwist
once larger applied forces begin to deform the molecule backbone. These apparently surprising findings have been interpreted in terms of a negative twist-stretch coupling which, in an elastic rod model,  causes the helix both to over-twist and to extend upon constant moderate stretching. Crucial to the authors' model is that, while the rod is stretched, the helix radius can vary and, precisely, it shrinks. Similar results have been obtained by magnetic tweezers experiments \cite{croq06} which find a linear \textit{extension versus twist} behavior, independent of sequence specificities and buffer conditions, in the low force domain.  

More generally, it has also been pointed out \cite{mad} that the theory of helical rods predicts different equilibrium twist conformations for a spring, subjected to a wrench. In particular, in the absence of an axial torque, the helical spring may first over-twist, then return to the initial state and eventually untwist as a function of the applied force.

However, while evidence has emerged concerning the interplay between twist and stretch \cite{wuite11} in the shaping of the DNA mechanics, a theoretical description of these phenomena at the molecular basis is still needed, mostly at forces for which DNA retains its double helical structure. Here, we address these issues by developing a mesoscopic Hamiltonian model for helicoidal molecules subject to external forces and compute their mechanical response to the stretching deformation. The focus is on very
short DNA helices which have recently attracted much interest in view of their remarkable flexibility \cite{archer,gole,tan15} despite the intrinsic inaccuracies encountered in the experimental characterization of the molecule structure at short length scales \cite{fenn,mastro,mazur}. At such scales, the DNA contour length may become even shorter than the persistence length therefore challenging the application of usual elastic rod models which treat the molecule as a continuum \cite{kimkim16}.

Our computational method for the double helix Hamiltonian is based on a path integral formalism \cite{io09,io10,io11} that incorporates in the base pair displacements those thermal fluctuations which are all the more relevant in short chains with large finite size effects.  
The same formalism has been used in some recent works \cite{io16b,io16a}, to study persistence length and cyclization probabilities of short DNA sequences with $\sim 100$ base pairs. Our analysis has shown that DNA chains maintain a large bendability also at short scales and may thus have persistence lengths smaller than those predicted by the worm-like-chain-model for kilo-base long molecules.  While these results point to the connection between global chain flexibility and inter-molecular interactions at the base pair level, it is true that the previously developed method assumed that the DNA sequences are torsionally constrained and therefore it cannot be applied to investigate the molecule twisting under a mechanical load. Accordingly, to pursue this task, we need to remove that constraint and formulate a more general computational program in order to evaluate the essential parameters which characterize the helix response to the external force, i.e. its diameter and the number of base pairs per helix turn.

The geometrical model for the helix is outlined in Section 2 while the Hamiltonian is presented in Section 3. The computational method is described in Section 4 and the results are shown in Section 5. Some conclusions are drawn in Section 6.

\section*{2. Model for the Helix }

We depict the backbone of a DNA molecule with $N$ base pairs, see Fig.~\ref{fig:1}(a), as a chain of $N-1$ segments connecting the points $O_i \,{}\, (i=\,1,...,N) $ which are equally spaced along the helix mid-axis. 

For short chains, the $O_i$'s can be considered as pinned to the sheet plane as the energetic cost to bend the backbone out of the plane is high. Accordingly, once the short chain closes into a loop, the supercoiling is mostly partitioned into twisting while the writhing contribution is essentially zero \cite{horo,fogg,irob}. 
Thus, in the absence of fluctuations, the planar molecule backbone is a freely jointed polygonal chain with the line segment $d$ being the rise distance between adjacent base pairs. 
For the i-th base pair, we define $x_{i}^{(1)}, \, x_{i}^{(2)}$ as the fluctuations of the two complementary mates measured from their respective  strands. 
With respect to $O_i$, the positions of the two pair mates are expressed respectively by, $r_{i}^{(1)}=\, -R_0/2 + x_{i}^{(1)}$ and $r_{i}^{(2)}=\, R_0/2 + x_{i}^{(2)}$, where $R_0$ is the separation between the complementary strands i.e., the equilibrium helix diameter.
It follows that the relative distance between the complementary bases, with respect to the helix mid-axis, is $r_{i}=\,r_{i}^{(2)} - r_{i}^{(1)}$.

Thus, the variable $r_{i}$ accounts for the radial base pair fluctuations around $R_0$. Note that $r_{i}$ may even become smaller than $R_0$ but the two bases cannot get too close to each other due to the electrostatic repulsion between negatively charged phosphate groups on complementary strands \cite{io11}.
The $O_{i}$'s also provide flexible hinges allowing adjacent displacements along the molecule stack, e.g. $r_{i}$ and $r_{i-1}$, to twist (
by the angle $\theta_i$) and to bend (by the angle $\phi_i$)  as shown in Fig.~\ref{fig:1}(b). Importantly, $\phi_i$ is also the angle between the helix diameters at adjacent sites $O_{i}$ and $O_{i-1}$ as, at any site, $r_{i}$ and $R_0$ are parallel.  While  $\phi_i$ and $\theta_i$ are integration variables in the computation of the partition function (as discussed below), both bending and twisting fluctuations are incorporated in our three dimensional helicoidal model. Thus, both angular and radial fluctuations contribute to determine the separation between adjacent base pairs, $\overline{d_{i,i-1}}$, measured by the distance $\overline{AB}$ in Fig.~\ref{fig:1}(b) and given by:

\begin{figure}
\includegraphics[height=8.0cm,width=8.0cm,angle=-90]{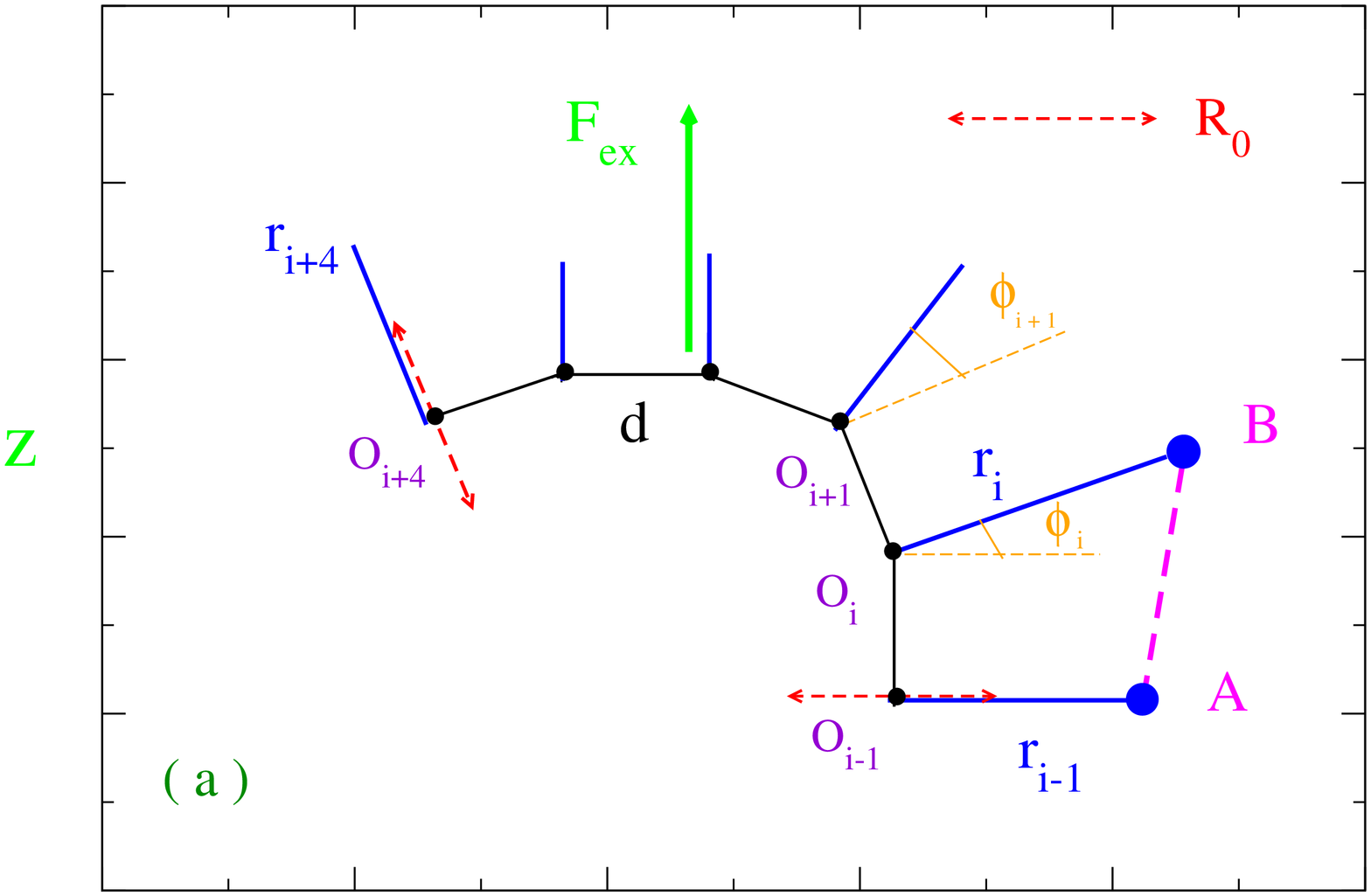}
\includegraphics[height=8.0cm,width=8.0cm,angle=-90]{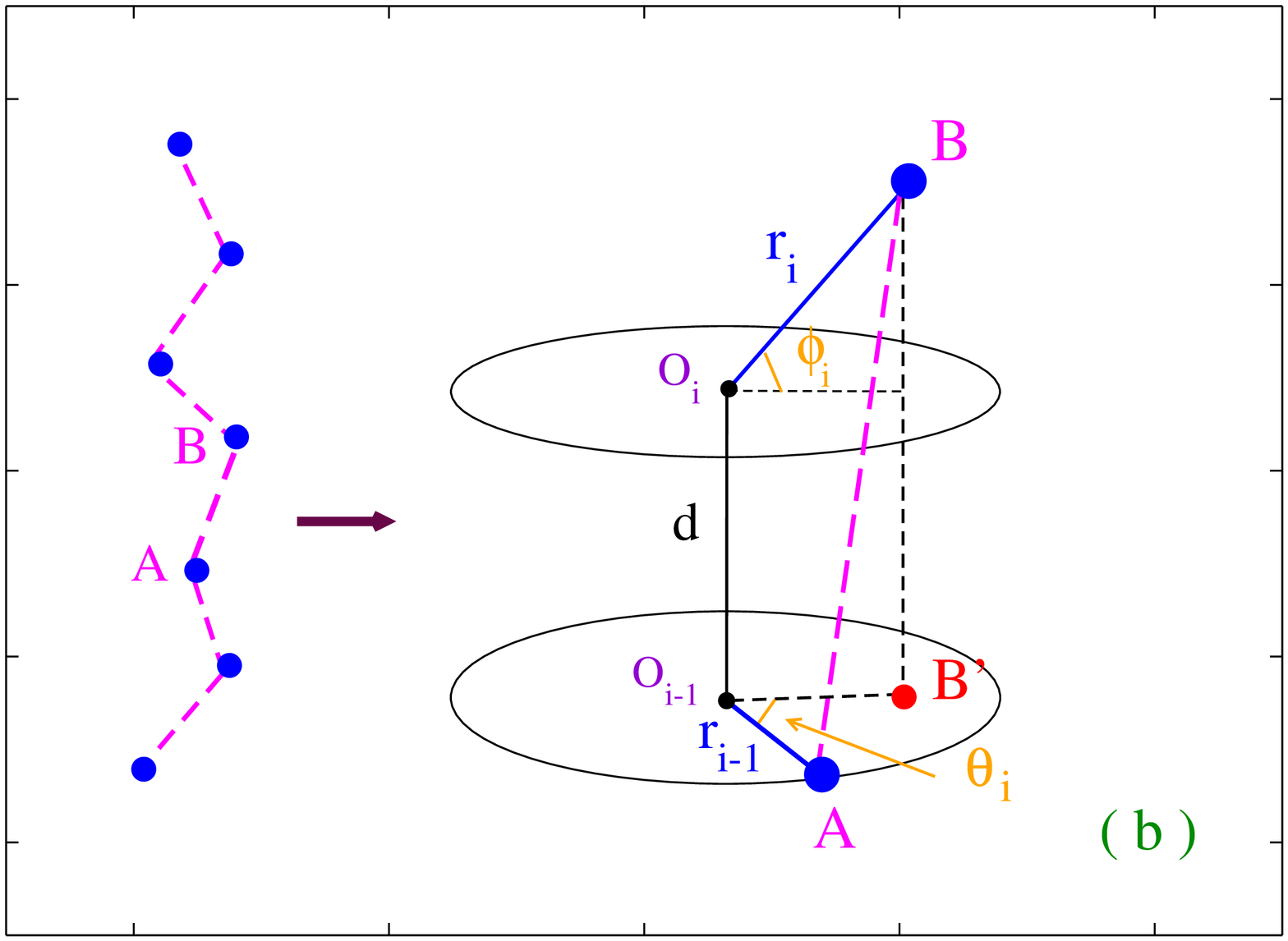}
\caption{\label{fig:1}(Color online)  
(a) {} Model for a chain of $N$ base pairs with external stretching force $F_{ex}$ applied along the $z-$ axis. $r_{i}$ is the relative distance between the two mates of the $i-th$ base pair. The $r_{i}$'s have variable amplitude and are measured with respect to the $O_i$'s which are arranged along the mid-axis of the helix (lying on the sheet plane). $R_0$ is the equilibrium helix diameter and $d$ is the rise distance between adjacent nucleotides along the molecule stack. In the absence of fluctuations, $r_{i}\equiv \,R_0$. Neighbor base pairs displacements, $r_{i}$ and $r_{i-1}$, are bent by the angle $\phi_i$. This is also the angle between the bond $\overline{O_{i+1}O_{i}}$ and $F_{ex}$ as the latter acts along the direction of the $\overline{O_{i}O_{i-1}}$ segment.
(b) {} Geometrical representation for the distance $\overline{AB}$ between base pairs displacements. Both bending and twisting fluctuations are incorporated in the stacking potential. $\theta_i$ is the torsional angle formed by adjacent $r_{i}$ and $r_{i-1}$. As defined in in Eq.~(\ref{eq:03}), $\theta_i$ is measured with respect to the average angle $<\theta_{i - 1}>$  of the preceding base pair along the stack. In the absence of bending, the model would reduce to a fixed-plane representation as depicted by the ovals.
}
\end{figure}

\begin{eqnarray}
& & \overline{AB}^2=\, \overline{BB'}^2 + \overline{AB'}^2 \, , \nonumber
\\
& & |\overline{BB'}| =\, d + r_i \sin \phi_i  \, , \nonumber
\\
& & \overline{AB'}^2=\, r_{i-1}^2 + (r_i \cos \phi_i)^2 -2 r_{i-1} \cdot r_i \cos \phi_i \cos \theta_i \, , 
\label{eq:0}
\end{eqnarray}

In the following, we set the values $R_0 \sim \,20 $\AA {} for the equilibrium helix diameter and  $d \sim \, 3$\AA {} for the bare rise distance. Small variations with respect to these values do not affect our results. Further features of the double stranded helix such as the presence of major and minor grooves and possible local distortions in helical parameters are not considered  here \cite{bates,orozco15}. 

Moreover, it is important to realize that our model essentially treats the unit consisting of a phosphate, a sugar and a base, i.e., the nucleotide, as a point-like object ignoring its internal degrees of freedom. While such coarse-grained level of description seems at this stage sufficient to capture the pattern of the twist-stretch dependence in short chains, we also mention that a finer level of resolution (looking at the internal structure of the nucleotide) has been achieved in computational studies of the elastic and melting properties of nucleic acids \cite{wang15}. 

The helicoidal chain is subject to an external force which is set along the direction of the rise distance $\overline{O_{i}O_{i-1}}$, as shown in Fig.~\ref{fig:1}(a), and assumed to act with the same modulus on any molecule segment. Note that the force direction is taken constant to comply with the experimental setup while the molecule backbone remains anchored to the sheet plane also after applying the external load.  In the presence of such tunable force field, we compute  \textit{i)} the average twisting fluctuations which allow to determine the number of base pairs per helix turn, i.e. the average helical repeat $< h_r >$, \textit{ii)} the average radial fluctuations $< r_i >$  which measure the molecule diameter  and \textit{iii)} the average extensions $< \overline{d_{i,i-1}} >$. All averages are taken over an ensemble of molecule configurations according to the method explained in Section 4.

\section*{3. Mesoscopic Model}

DNA in solution is stable at physiological temperatures despite the energetic molecular bumping due to the environment. In fact, as the room temperature thermal energy ($k_B T \sim 25 meV$) is comparable to the hydrogen bond base pair energy \cite{nina10}, fluctuational bubbles can form spontaneously and transiently along the helix \cite{gueron,benham,bonnet,russu,rapti,metz09,bishop09} whereas the molecule strands remain intact due to the overall strong covalent bonds acting between adjacent nucleotides \cite{kame06}. Random coil molecule conformations are favored by entropic forces which bend the intra-strand stacking bonds although DNA becomes stiff at scales of order of the typical persistence length, i.e. $\sim 50 nm$. Such length may be significantly smaller for short DNA \cite{porsch,menon,manghi15} and  sequences with $ \sim 100$ base pairs or less could maintain an intrinsic flexibility which favors the molecule looping and the loop stability against fluctuational forces \cite{ejte15,kim16}.

To account for the main interactions that stabilize the helix we use a mesoscopic Hamiltonian containing a one-particle base pair potential $V_{1}[r_i]$ and a two-particle stacking term $V_{2}[ r_i, r_{i-1}, \phi_i, \theta_i]$ also dependent on the angular degrees of freedom. Their analytic form reads:

\begin{eqnarray}
& &V_{1}[r_i]=\, V_{M}[r_i] + V_{Sol}[r_i] \, , \nonumber
\\
& &V_{M}[r_i]=\, D_i \bigl[\exp(-b_i (|r_i| - R_0)) - 1 \bigr]^2  \, , \nonumber
\\
& &V_{Sol}[r_i]=\, - D_i f_s \bigl(\tanh((|r_i| - R_0)/ l_s) - 1 \bigr) \, , \nonumber
\\
& &V_{2}[ r_i, r_{i-1}, \phi_i, \theta_i]=\, K_S \cdot \bigl(1 + G_{i, i-1}\bigr) \cdot \overline{d_{i,i-1}}^2  \, , \nonumber
\\
& &G_{i, i-1}= \, \rho_{i, i-1}\exp\bigl[-\alpha_{i, i-1}(|r_i| + |r_{i-1}| - 2R_0)\bigr]  \, . \nonumber
\\ 
\label{eq:00}
\end{eqnarray}

$V_{1}[r_i]$ is the sum of two terms:\textit{ 1)} a hydrogen bond Morse potential $( V_{M}[r_i] )$ for the $i-th$ base pair with dissociation energy $D_i$ and spatial range measured by $b_i$. \textit{2)} A solvent term $( V_{Sol}[r_i] )$ which modifies the Morse plateau through the parameters $f_s$ and $l_s$.   The one-particle potential has been used in the past \cite{proho,joy09,weber09} to simulate the denaturation of the double stranded DNA occurring for base pair separation large enough to sample the Morse plateau. The same analytic form of $V_{1}[r_i]$ is suitable to model thermally stable hydrogen bonds \cite{baird} as in the current analysis provided that the base pair fluctuations remain smaller than the helix radius.

As mentioned above, the base pair separation may become shorter than $R_0$ but too large helix contraction are discouraged by the large energetic cost. This physical requirement is implemented in the code by discarding those paths for which the repulsive energy becomes too large, i.e. larger than the dissociation energy: $V_{M}[r_i] \geq D_i$. This implies that the base pair amplitudes must fulfill the condition, $|r_i| - R_0 < - \ln(2 / b_i)$.

The stacking potential $V_{2}[ r_i, r_{i-1}, \phi_i, \theta_i]$
contains an elastic force constant $ K_S$ and  nonlinear contributions weighed by the parameters $\rho_{i, i-1}$, $\alpha_{i, i-1}$ as originally proposed in the context of a simple ladder model for DNA \cite{pey2}. $V_{2}$ depends on the angular degrees of freedom through the fluctuating distance $\overline{d_{i,i-1}}$ in Eq.~(\ref{eq:0}). The physical motivations underlying the choice of these potentials have been widely discussed in previous works \cite{io11,io12} to which we refer for details.  The model parameters, taken hereafter as in \cite{io16b,io16a}, are appropriate to homogeneous sequences of $GC$- base pairs which have sizeable dissociation energies and are stable against thermal fluctuations. 
Certainly quantitative predictions of the twist-stretch relations may be derived for specific heterogeneous sequences since, in general, the helical properties are  dependent on the base sequences, as also recently pointed out in the different context of charge transfer in B-DNA segments \cite{sims15,sims16}.

With this caveat, our helicoidal molecule with $N$ base pairs of reduced mass $\mu$, stretched by a force $F_{ex}$ as in Fig.~\ref{fig:1}(a), is described by the Hamiltonian:

\begin{eqnarray}
& &H =\, H_a[r_1] + \sum_{i=2}^{N} H_b[r_i, r_{i-1}, \phi_i, \theta_i] \, , \nonumber
\\
& &H_a[r_1] =\, \frac{\mu}{2} \dot{r}_1^2 + V_{1}[r_1] \, , \nonumber
\\
& &H_b[r_i, r_{i-1}, \phi_i, \theta_i]= \,  \frac{\mu}{2} \dot{r}_i^2 + V_{1}[r_i] + V_{2}[ r_i, r_{i-1}, \phi_i, \theta_i] - F_{ex} d  \cos\bigl( \sum_{k=1}^{i-1}\phi_k \bigr)  \, . \nonumber
\\ 
\label{eq:01}
\end{eqnarray}

$H_a[r_1]$ is taken out of the sum as the first base pair has no preceding neighbor along the molecule backbone.
Path integral techniques can be applied to Eq.~(\ref{eq:01}) to derive the partition function $Z_N$  following a well-tried method \cite{io14b}. The latter assumes that the base pair distance $r_i$ in Eq.~(\ref{eq:01}) is a  trajectory $r_i(\tau)$ depending on the imaginary time $(\tau)$ defined as the analytic continuation of the real time to the imaginary axis \cite{fehi}.
Accordingly $r_i(\tau)$ is expanded in  Fourier series whose coefficients generate an ensemble of possible base pair trajectories. As this is done for any base pair in the chain, the technique permits building an ensemble of molecule configurations which statistically contribute to the path integral with a Boltzmann weight.  Technically, the size of the ensemble, i.e., the number of trajectories, is increased until numerical convergence in $Z_N$ is achieved. This condition corresponds to the state of thermodynamic equilibrium for the system. Including also fluctuations over the bending and twisting degrees of freedom, with their respective cutoffs, the general expression for $Z_N$  reads:

\begin{eqnarray}
& &Z_N=\, \oint Dr_{1} \exp \bigl[- A_a[r_1] \bigr]   \prod_{i=2}^{N}  \int_{- \phi_{M} }^{\phi_{M} } d \phi_i \int_{- \theta_{M} }^{\theta _{M} } d \theta_{i} \oint Dr_{i}  \exp \bigl[- A_b [r_i, r_{i-1}, \phi_i, \theta_i] \bigr] \, , \nonumber
\\
& &A_a[r_1]= \,  \int_{0}^{\beta} d\tau H_a[r_1(\tau)] \, , \nonumber
\\
& &A_b[r_i, r_{i-1}, \phi_i, \theta_i]= \,  \int_{0}^{\beta} d\tau H_b[r_i(\tau), r_{i-1}(\tau), \phi_i, \theta_i] \, ,
\label{eq:02}
\end{eqnarray}

where $\beta$ is the inverse temperature. $\oint {D}r_i$ is the measure of integration over the space of the Fourier coefficients whose temperature dependent cutoffs are consistently defined in the path integral method \cite{io11a,io03}. While this peculiar feature makes the method also suitable to study thermally driven effects such as bubble formation and DNA denaturation \cite{io11,io13},  the calculations hereafter presented are carried out at room temperature. From Eq.~(\ref{eq:02}), the free energy of the system is computed as: $F=\, -\beta ^{-1} \ln Z_N$.

\section*{4. Computational Method}

The maximum amplitude of the bending fluctuations in Eq.~(\ref{eq:02}) is set by $\phi_{M}$ which at a first stage is taken constant, i.e., $\phi_{M}=\,\pi /2$. The latter choice allows for kinks formation as a possible mechanism, first put forward long time ago \cite{crick}, to reduce the bending energy between adjacent base pairs \cite{zocchi13}. This has also been suggested more recently by molecular dynamics simulations \cite{harris} and analysis of (un)looping probabilities in short chains based on single molecule fluorescence resonance energy transfer \cite{kim14}. While the polymer flexibility can be directly related to the average bending angles between any two distant monomers \cite{soder,volo10}, computation of persistence length has shown that linear short sequences have an intrinsic flexibility mainly ascribed to their terminal base pairs \cite{io16a,weber15}.

The novelty brought about here lies in the way twisting fluctuations are treated for the model drawn in Fig.~\ref{fig:1}(b). Our idea is that the twist variable $\theta_{i}$ in Eq.~(\ref{eq:02}) should be measured with respect to the ensemble averaged $<\theta_{i - 1}>$ value obtained for the preceding base pair along the stack. Such value is incremented by $2\pi / h_r$, where the number of base pairs per helix turn is taken as an input parameter to be chosen within an appropriate range. For any assumed $h_r$, we further 
admit there may be a fluctuational range around the value $\, <\theta_{i - 1}>  + 2\pi / h_r \,$,  with $\theta_{i}^{fl}$ being the twist fluctuation integration variable.
Formally our Ansatz is expressed by:

\begin{eqnarray}
& &\theta_i =\, <\theta_{i - 1}>  + 2\pi / h_r + \theta_{i}^{fl} ,  \nonumber
\\
& &h_r \in \, [h_r^{min}, \, h_r^{max}]  ,  \nonumber
\\
& &h_r^{max} - h_r^{min}=\, n \cdot \Delta h_r \, \, {} \hskip 2cm n \in \mathbb{Z} \, ,
\label{eq:03}
\end{eqnarray}

where $\Delta h_r$ is the increment in the helical repeat range. The accuracy of the method grows with the number $n$ of values sampled in such interval. For kilo-base B-DNA in solution the experimental helical repeat under physiological condition is, $h_r^{exp}=\, 10.4$ \cite{wang}, whereas a smaller value $h_r\sim \, 10$ has been estimated for DNA wrapped around histone proteins \cite{levitt78,hayes}.  In general, thermal fluctuations tend to untwist the helix \cite{duguet}. In short chains the helical repeat may  differ from $h_r^{exp}$ and certainly it changes under the action of applied forces.   Accordingly, our strategy consists of assuming a broad range around $h_r^{exp}$, \, e.g., $h_r^{min}=\,6$,\, $h_r^{max}=\,14$, with a coarse partition ($\Delta h_r=\,0.125$) and searching for the energetically most convenient helical conformations as a function of the external force. Next we take a finer partition of the range (enhancing both $n$ and the CPU time) and repeat the search until the selected helical conformations are stable against $\Delta h_r$.

Central to this work is the calculation of the average twist angles formally given by:

\begin{eqnarray}
& &< \theta_i >_{(i \geq 2)} =\,  < \theta_{i-1} > + \frac{2\pi}{h_r}  + \frac{\int_{-\theta_{M}}^{\theta_{M}} d \theta_{i}^{fl} \cdot ( \theta_i^{fl} ) \int_{- \phi _{M}}^{\phi _{M}} d \phi_i  \oint Dr_{i} \exp \bigl[- A_b [r_i, r_{i-1}, \phi_i, \theta_i]  \bigr]}{ \int_{-\theta_{M}}^{ \theta_{M}} d \theta_{i}^{fl} \int_{- \phi _{M} }^{\phi _{M} } d \phi_i  \oint Dr_{i} \exp \bigl[- A_b [r_i, r_{i-1}, \phi_i, \theta_i]  \bigr]} \, . \nonumber
\\
\label{eq:04}
\end{eqnarray}

Numerical convergence on the average twist angles is achieved by setting the cutoff on the twist fluctuations, $\theta_{M}=\,\pi /4$. For the first base pair in the chain, $\theta_1 \equiv \, \theta_{1}^{fl}$  hence,   $< \theta_1 >=\,0$.

Then, from Eq.~(\ref{eq:04}), we derive the average helical repeat:

\begin{eqnarray}
< h_r >=\,\frac{2\pi N}{< \theta_N >} \, .
\label{eq:05}
\end{eqnarray}

Computing Eqs.~(\ref{eq:04}),~(\ref{eq:05}) for any $h_r$ in Eq.~(\ref{eq:03}), the program generates a set of values $\{ < h_r >_{j} \, ,  \, (j=1,...,n) \} $ which, in general,  deviate from the initially chosen $h_r$-values of our simulation.
After obtaining such set, we select by free energy minimization the equilibrium value $ < h_r >_{j^{*}} $. As this procedure is iterated for any applied force, the twist-stretch profile for a specific molecule is eventually derived.  

Next,  integrations over the ensemble of base pair configurations are performed to get: \textit{1)} the average base pair radial fluctuations which determine the average helix diameter, $R_0 + < R >$ \cite{note} , 

\begin{eqnarray}
& &< r_i > =\,  \frac{\oint Dr_{i} \cdot r_i \int_{-\theta_{M}}^{\theta_{M}} d \theta_{i}^{fl}   \int_{-\phi _{M}}^{\phi _{M}} d \phi_i   \exp \bigl[- A_b [r_i, r_{i-1}, \phi_i, \theta_i]  \bigr]}{ \int_{-\theta_{M}}^{\theta_{M}} d \theta_{i}^{fl} \int_{-\phi _{M} }^{\phi _{M} } d \phi_i  \oint Dr_{i}  \exp \bigl[- A_b [r_i, r_{i-1}, \phi_i, \theta_i]  \bigr]} \, , \nonumber
\\
& &< R >=\frac{1}{N} \sum _{i=1}^{N} < r_i > \, 
\label{eq:06}
\end{eqnarray}

and \textit{2)} the average distances between base pairs along the stack,

\begin{eqnarray}
& &< \overline{d_{i,i-1}} > =\,  \frac{\oint Dr_{i}  \int_{-\theta_{M}}^{\theta_{M}} d \theta_{i}^{fl}   \int_{ -\phi _{M}}^{\phi _{M}} d \phi_i  \cdot \overline{d_{i,i-1}} \exp \bigl[- A_b [r_i, r_{i-1}, \phi_i, \theta_i]  \bigr]}{ \int_{-\theta_{M}}^{\theta_{M}} d \theta_{i}^{fl} \int_{-\phi _{M} }^{\phi _{M} } d \phi_i  \oint Dr_{i}  \exp \bigl[- A_b [r_i, r_{i-1}, \phi_i, \theta_i]  \bigr]} \, , \nonumber
\\
& &< d >=\frac{1}{N-1} \sum _{i=2}^{N} < \overline{d_{i,i-1}} > \, . 
\label{eq:07}
\end{eqnarray}

From Eq.~(\ref{eq:07}), one finally estimates the force induced extension per base pair with respect to the un-stretched base pair separation:

\begin{eqnarray}
& &\Delta z =\, \frac{< d (F_{ex}) >}{< d (0) >} - 1 \, \, \, . 
\label{eq:08}
\end{eqnarray}

The $< R >$'s and $\Delta z$'s have been calculated for any helical repeat in Eq.~(\ref{eq:03}) but the values reported hereafter are those which correspond to the equilibrium conformation $ < h_r >_{j^{*}} $ selected by the free energy minimum.

\section*{5. Results}

\begin{figure}
\includegraphics[height=12.0cm,width=12.0cm,angle=-90]{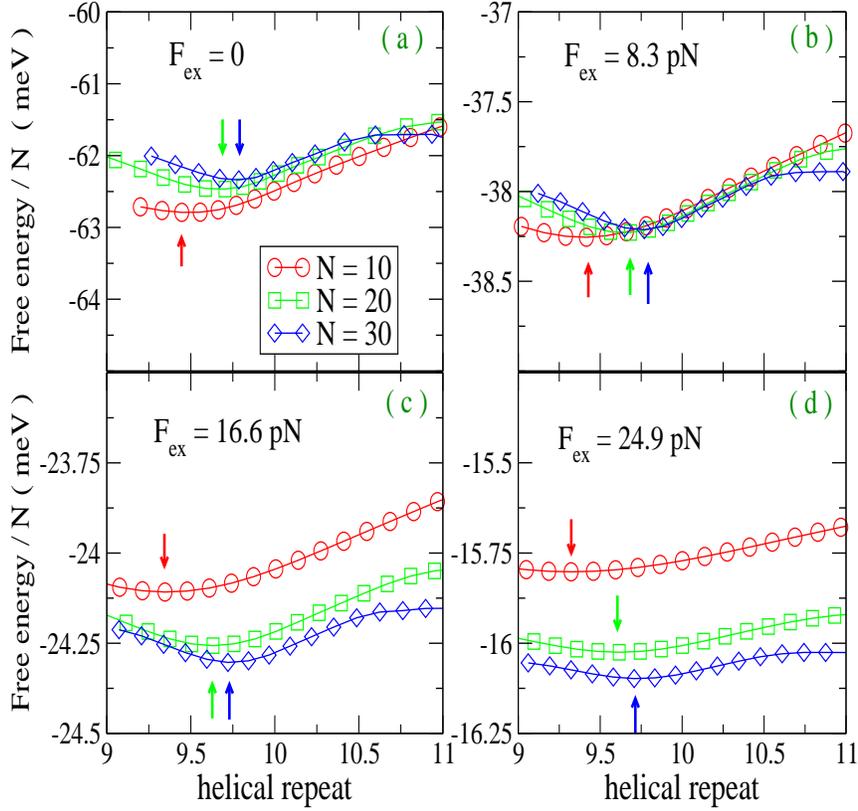}
\caption{\label{fig:2}(Color online) 
Free energies per base pair, calculated for three homogeneous sequences,  as a function of the average helical repeat given by Eq.~(\ref{eq:05}). The room temperature free energies are computed over a range of helical repeat values with incremental step $\Delta h_r =\, 0.125$ (see text).  The four panels refer to the zero force case and three values of applied force. The arrows mark, for each sequence, the free energy minimum. 
}
\end{figure}

Our method is illustrated in Fig.~\ref{fig:2} where the free energy per base pair ($F / N$) is plotted versus $< h_r >$ for three short homogeneous sequences, both in the absence of force and for three increasing $F_{ex}$ strengths. An increment $\Delta h_r=\,0.125$ has been assumed in these calculations.

Given our choice of model parameters, the free energy scale for $F_{ex}=\,0$ (panel (a) ) shows that $F / N$ is in the range of the experimental values, i.e., $1 - 1.5 \, kcal/mol$ \cite{kame06}. This holds for all sequences. Applied forces have the general effect to straighten the coiled molecules thus reducing the entropy. It follows that $F / N$ should become larger by enhancing $F_{ex}$.  Consistently the free energy scales display growing (less negative) values in the three successive panels. Also note that $F / N$ increases more rapidly for shorter molecules and, for sizeable $F_{ex}$, it gets distinctly higher for the $N=\,10$ molecule as shown in panels (c), (d).  This indicates that external forces more effectively straighten the bonds in shorter molecules.
The arrows mark, for each sequence, the free energy minima occurring at the above defined $ < h_r >_{j^{*}} $. Such values shift downwards (towards smaller helical repeats) by increasing $F_{ex}$ from (a) to (d), pointing to a general helix over-twisting as a consequence of the mechanical stretching.

\begin{figure}
\includegraphics[height=8.0cm,width=8.0cm,angle=-90]{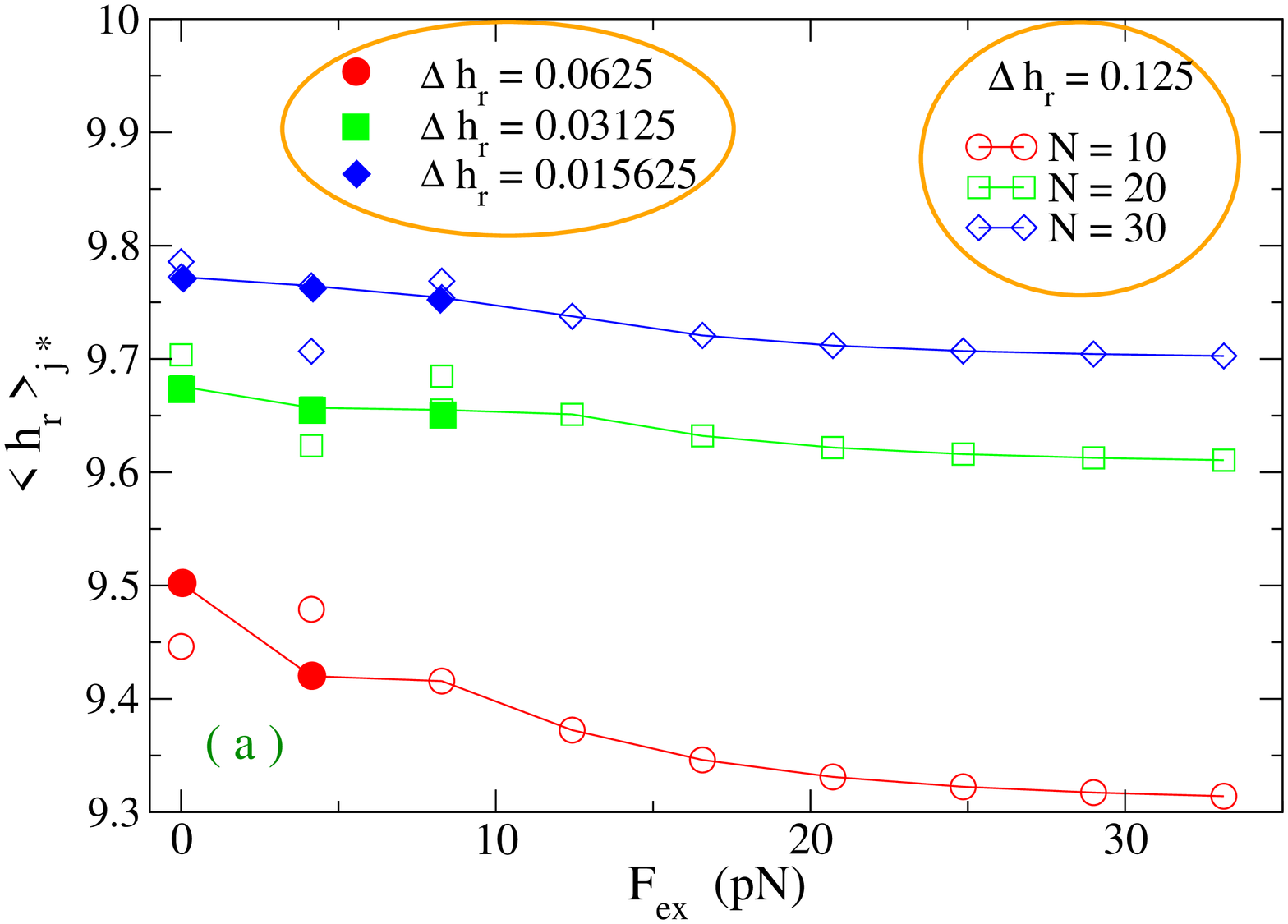}
\includegraphics[height=8.0cm,width=8.0cm,angle=-90]{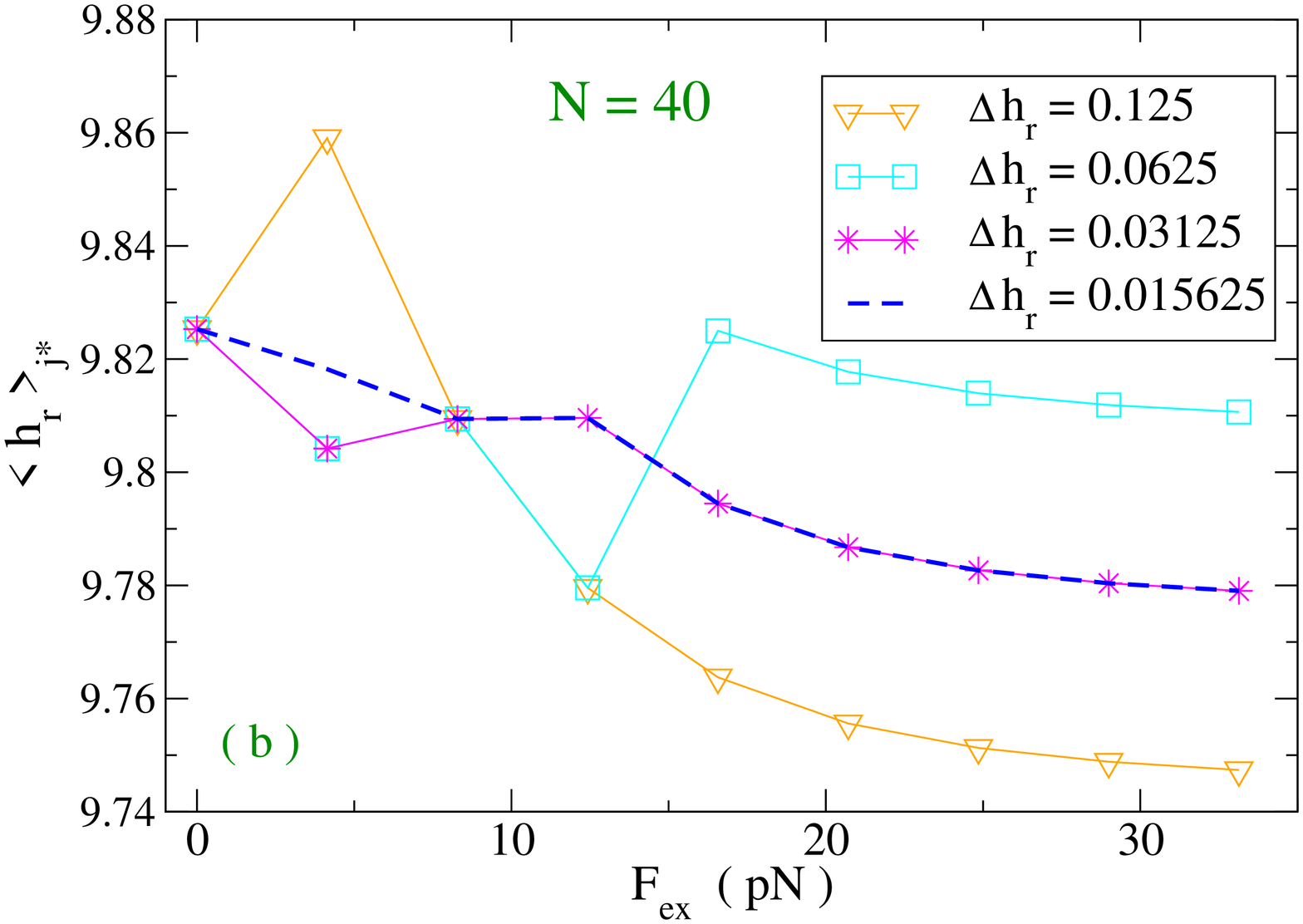}
\caption{\label{fig:3}(Color online) 
(a) {} Ensemble averaged helical repeats as a function of the external force, calculated via Eqs.~(\ref{eq:04}),~(\ref{eq:05}) for three sequences. The empty symbols refer to a coarse partition in Eq.~(\ref{eq:03}). In the low forces regime, finer partitions (smaller $\Delta h_r$) are required to obtain numerical convergence.
(b) {} As in (a) but for a longer sequence. The average helical repeats are computed by tuning $\Delta h_r$. 
}
\end{figure}

These findings are made more evident in Figs.~\ref{fig:3} where the $ < h_r >_{j^{*}} $ are directly reported as a function of the mechanical force. For the three sequences, $N=\,10 \, , \,20 \, , 30$,  in Fig.~\ref{fig:3}(a),  the computation is first carried out with the same coarse partition of the helical repeat range as in Fig.~\ref{fig:2}.

It turns out that the $ < h_r >_{j^{*}} $'s  generally decrease versus $F_{ex}$ but, in the low forces regime, the plots display a non-monotonic behavior (empty symbols). 
However, reducing $\Delta h_r$, a monotonic dependence on $F_{ex}$ is recovered for all sequences as visualized by the filled symbols. Only the $ < h_r >_{j^{*}} $'s calculated  for $F_{ex} \sim 4, \, 8$ pN are reported in the figure as, at higher forces, the $ < h_r >_{j^{*}} $'s do not substantially change by taking a finer mesh. In general, a finer partition of the range in Eq.~(\ref{eq:03}) should be used to achieve numerical convergence in the low force regime in which the helical repeat has a larger gradient. Instead, applying larger forces which significantly stretch the molecule,  the $ < h_r >_{j^{*}} $'s become more stable and the transitions among twist conformations with different energy are reduced. Hence, convergence is already achieved using a coarser $\Delta h_r$ mesh.

Increasing $F_{ex}$ has the effect to dampen the thermal fluctuations and set the molecules in a more ordered conformation. This explains why the search for the equilibrium value $ < h_r >_{j^{*}} $ can be made with a coarser resolution at larger applied loads. 
It is also found that, for longer sequences,  smaller $\Delta h_r$'s are necessary in order to determine with accuracy the $ < h_r >_{j^{*}} $'s  at low external forces.   

The interplay between sequence length and $\Delta h_r$ is further pointed out in Fig.~\ref{fig:3}(b) for a $N=\,40$ sequence. In this case, also at large applied forces,  the computation converges only with a fine partition,  $\Delta h_r=\,0.03125$, while a finer step, $\Delta h_r=\,0.015625$, is required to get convergent and monotonic behavior at small $F_{ex}$. Summing up, the relevant feature put forward by our plots is that short DNA sequences over-twist under mechanical stretching and these structural deformations appear as a response dictated by energetic convenience. While this result confirms the trend suggested by the measurements of ref.\cite{busta06} albeit on kilo-base long molecules, it would be interesting to test whether chains of a few tens of base pairs behave in a similar way.  Besides the length, also sequence specificities may affect mechanical properties and twisting flexibility in heterogeneous chains \cite{peters13}. 

Interestingly however, from Figs.~\ref{fig:3}, one can also estimate the force induced change in superhelical density $\sigma$. For sequences with planar molecular axis, $\sigma \equiv \, (Tw - Tw_{0}) / Tw_{0}$, where $Tw_{0}$ is the twist number in the absence of forces. Looking for instance at the plot for $N=\,30$ and taking the computed $ < h_r >_{j^{*}} $ at $F_{ex} \sim 20$ pN, we obtain $\sigma \sim 0.0076$ which is in fair agreement with the experimental  $\sigma \sim 0.01$ at $F_{ex} \sim 18$ pN \cite{busta06} measured in long sequences.

\begin{figure}
\includegraphics[height=8.0cm,width=8.0cm,angle=-90]{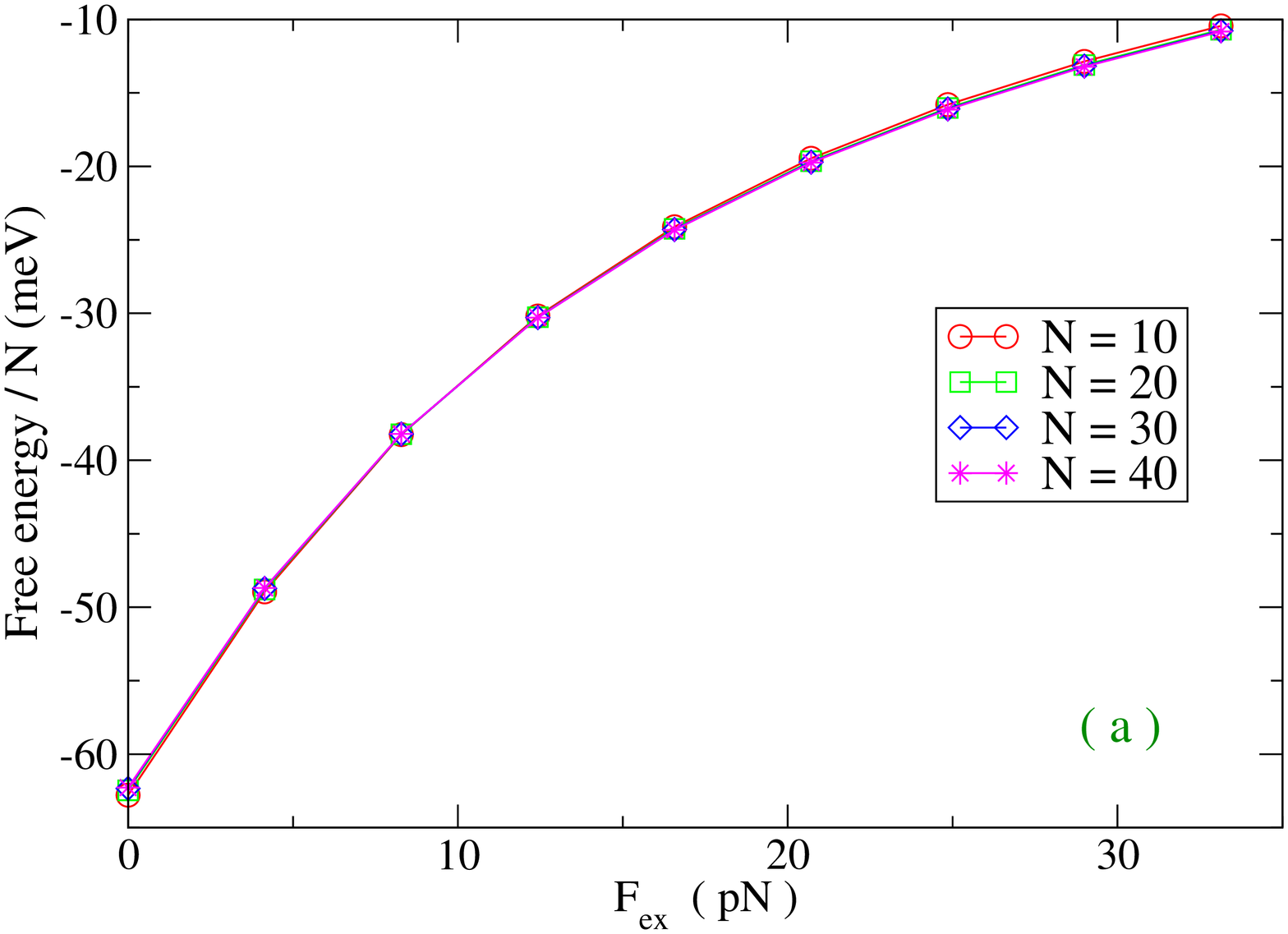}
\includegraphics[height=8.0cm,width=8.0cm,angle=-90]{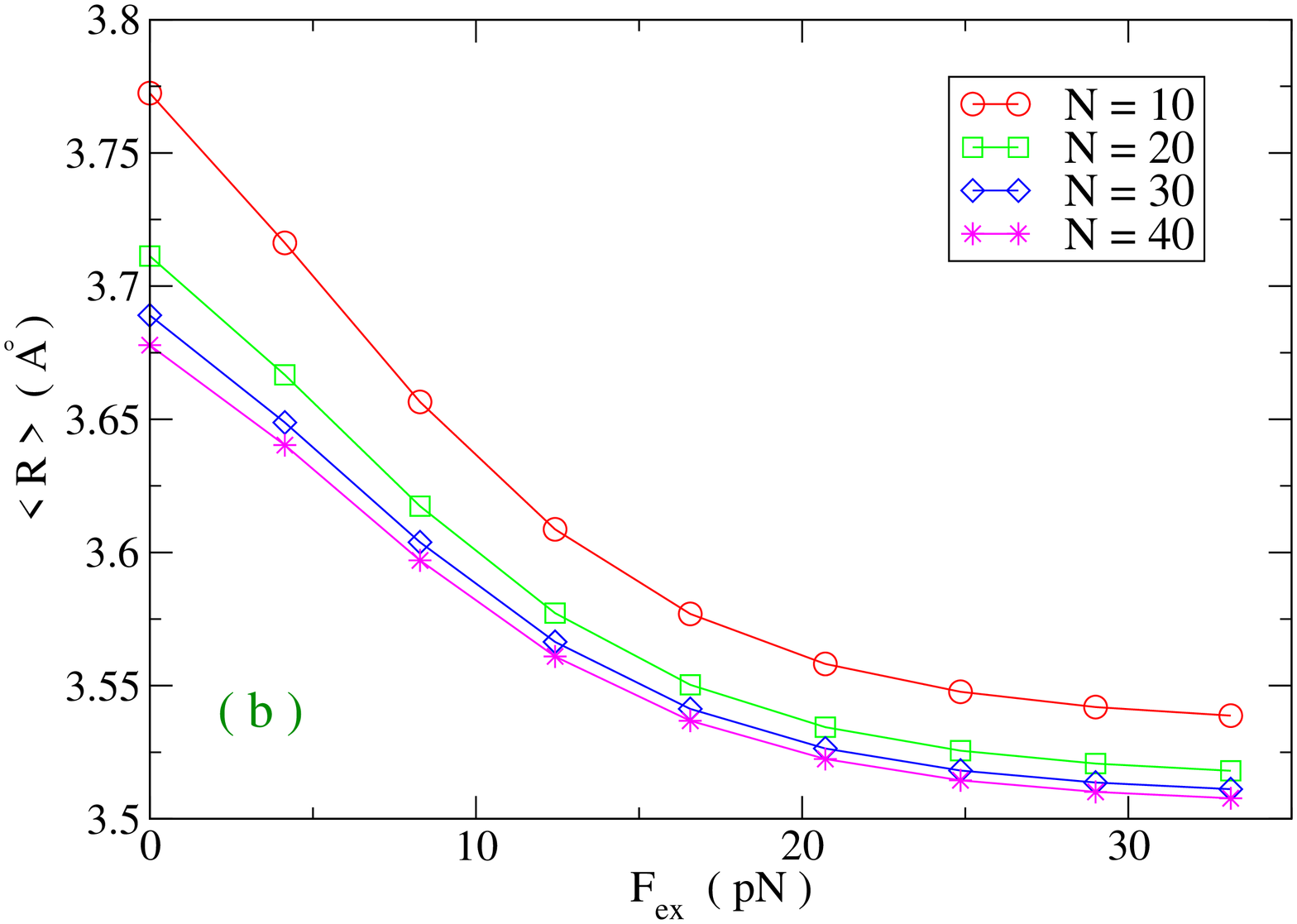}
\caption{\label{fig:4}(Color online)
(a) {} Room temperature free energies per base pair, calculated for four sequences,  as a function of the external force. The free energy per base pair values correspond to the minima indicated by arrows in Fig.~\ref{fig:2}. Note that such values essentially overlap for various $N$ due to the enlarged energy scale taken here respect to Fig.~\ref{fig:2}.
(b) {} Ensemble averaged base pair fluctuations, calculated via Eq.~(\ref{eq:06}), as a function of the external force for four sequences.
}
\end{figure}

Physically, the shrinking of the helix diameter could be associated to the shortening of the radial base pair fluctuations induced by the external force.  All-atom simulations of stretched and compressed DNA structures have previously suggested that helix diameter reduction and over-twisting are concomitant effects \cite{olson99}. 
We address this point in Figs.~\ref{fig:4} which display for the four short chains: (a) the free energy per base pair calculated, via Eq.~(\ref{eq:02}), at the helical equilibrium values $ < h_r >_{j^{*}} $ and (b) the average radial fluctuations, superimposed to the diameter $R_0$, obtained from Eq.~(\ref{eq:06}).
As noticed above, the free energies rapidly increase under the ordering effect of the applied load and, on this scale, the plots for the four chains essentially overlap. However, the growth rate of $F / N$  decreases versus $F_{ex}$ and, markedly, above $\sim 20$ pN. 
Likewise, once $F_{ex}$ is applied,  $< R >$  quickly drops from its value at zero force  and, for all chains, tends to saturate above $\sim 20$ pN. Nevertheless, intrinsic base pair fluctuations persist also in the presence of large external loads. Then, we find that the DNA over-twisting and the reduction of the helix diameter are indeed correlated structural properties also at the very short length scales considered in this work. Molecular models based on energy minimization have suggested that the shrinking of the DNA diameter could be associated to a negative inclination of the base pairs towards the minor groove \cite{croq06,zachar15}.

\begin{figure}
\includegraphics[height=12.0cm,width=12.0cm,angle=-90]{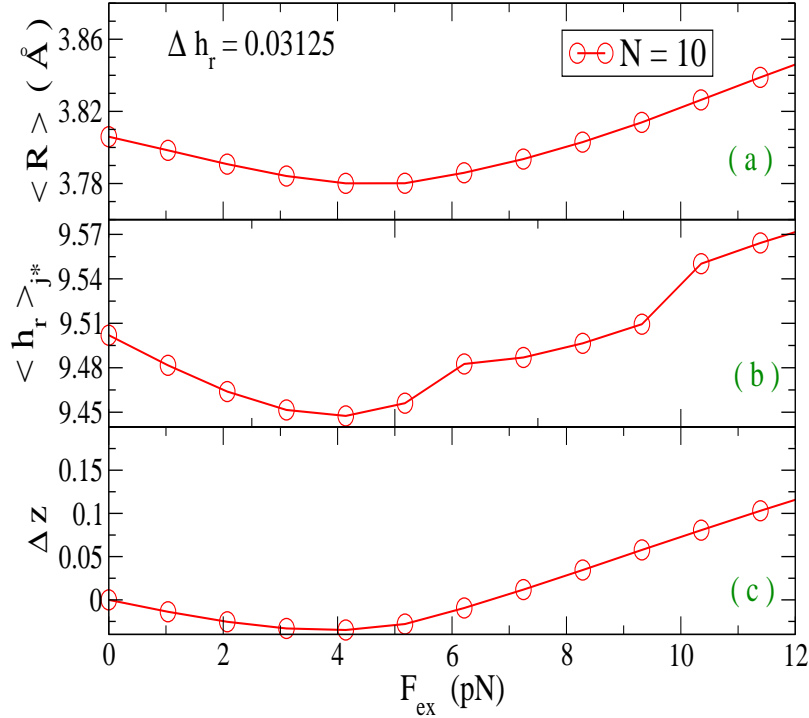}
\caption{\label{fig:5}(Color online) 
(a) {} Average radial fluctuations calculated from Eq.~(\ref{eq:06}) versus the external force $F_{ex}$;
(b) {} Average helical repeat obtained from Eqs.~(\ref{eq:04}),~(\ref{eq:05}) after minimizing the free energy for any $F_{ex}$;
(c) {} Average intra-strand extension per base pair from Eqs.~(\ref{eq:07}),~(\ref{eq:08}).  Both (a) and (c)  are calculated for the equilibrium helical conformations $ < h_r >_{j^{*}} $ in (b). The amplitude of the bending fluctuations is assumed to be dependent on $F_{ex}$ as described in the text.
}
\end{figure}

As mentioned in the Introduction, experiments indicate that stretched kilo-base long DNA first over-twists and eventually untwists if the applied loads exceeds $\sim 30$ pN.
Instead, the results presented so far do not account for the helix untwisting at large $F_{ex}$ as the $ < h_r >_{j^{*}} $'s decrease monotonically versus $F_{ex}$ in Figs.~\ref{fig:3}. The source of this discrepancy lies however in the way we have treated the bending fluctuations and precisely their maximum amplitude $\phi_{M}$ at the beginning of Section 4. Once the molecule is stretched, the intra-strand bonds straighten and therefore it is likely that the amplitude of the bending fluctuations between adjacent nucleotides is reduced
whereas there is no physical reason to impose a similar constraint on the twisting fluctuations which may remain large also for sizeable external loads. Accordingly, a consistent model should contemplate a force dependent integration cutoff $\phi_{M}(F_{ex})$. Although we are not aware of experiments providing data to which we may fit such function, some plausible functional forms have been guessed to test their effects on the output of our numerical code. Taking for instance,

$\phi_{M}(F_{ex})=\, {\pi }[ 1 - (c \cdot  F_{ex})^2 ] / 2$ \,  with \, $c^{-1}=\, 24$ pN, we obtain the results displayed in Fig.~\ref{fig:5} for a $N=\,10$  sequence. Similar plots are obtained for longer chains. The computation follows the same pathway described above. The transition between over-twisting and untwisting regime is now found and it occurs at $\sim 4$ pN.  Consistently, the average radial fluctuations (panel (a)) decrease up to $\sim 4$ pN and expand once the helix untwists under larger loads. While the latter force value is only indicative and clearly depends on the specific $\phi_{M}(F_{ex})$, there is no cogent reason why the transition at $\sim 30$ pN, observed for kilo-base sequences
\cite{busta06}, should take place also for short sequences in the same force regime. 
In this regard, experimental research may clarify the behavior of short DNA under stretching and foster further theoretical investigation.
However, the fact that the helix diameter and twisting conformations display a non-monotonic behavior versus the applied force corroborates our choice to impose a force dependent cutoff only on the bending fluctuations.

Likewise the intra-strand base pair separation (panel (c)), calculated via Eqs.~(\ref{eq:07}),~(\ref{eq:08}), slightly contracts with respect to the zero force conformation in the over-twisting regime and then stretches above $\sim 4$ pN. Thus we find that, under the action of a \textit{variable force}, the end-to-end distance grows when the helix untwists and, vice-versa, it shortens when the helix over-twists. While this pattern seems physically intuitive, it is only apparently at variance with ref.\cite{busta06} as in the latter the experimental setup was such that the changes in DNA extension due to an imposed over-twist had been measured at \textit{constant force}. 

To test whether our program reproduces the experimental trend also on this specific point, we should assume a constant force, say $F_{ex}^{'}$, and study the average base pair distance \, $< d >$ \, as a function of $< h_r >$. 
The experimentally imposed over-twist is simulated by 
considering the $< h_r >$'s which are smaller than the value $ < h_r >_{j^{*}} $ for the equilibrium conformation in the presence of $F_{ex}^{'}$. Note that such set of $< h_r >$'s has been computed \textit{before} selecting $ < h_r >_{j^{*}} $ by free energy minimization.
Clearly, any assumed $< h_r >$ differing from $ < h_r >_{j^{*}} $ corresponds to a condition in which the system is away from its free energy minimum. 
In this case we find (not shown here) that $< d >$ grows by reducing $< h_r >$ hence, the helical extension increases upon over-twisting consistently with the measurements. At the same time  $< R >$ decreases by reducing $< h_r >$. Thus, short dsDNA is predicted to shrink and extend when over-twisted under a constant load. Further quantitative analysis of the twist-extension relation may be performed on the base of models which account for the sequence specificities of the double helical structure.

\section*{6. Conclusions }

We have developed a method to evaluate the deformations of the DNA helical structure induced by an external load whose strength is such to compete on the energy scale with the base pair thermal fluctuations. Applied forces in the range of a few pico-Newton may in fact straighten the intra-strand molecule bonds and dampen the amplitudes of the base pair radial distances. Motivated  by ongoing research on the properties of short DNA helices, we have analyzed the interplay between helix stretching and twisting using a mesoscopic Hamiltonian model which sees the double stranded molecule as a sequence of interacting base pairs. This discrete approach seems particularly suitable to DNA molecules whose contour length may be shorter that their persistence length. In fact, at such short scales, the applicability of continuum elastic rod models has been questioned over the last years. The statistical mechanics of the mesoscopic Hamiltonian has been formulated by a well-established finite temperature path integration method which treats the inter-strand and intra-strand interactions in terms of the ensemble of trajectories for the base pair radial configurations. Also bending and twisting fluctuations between neighboring base pairs along the molecule backbone are included in the partition function.

Introducing a computational scheme which allows the molecule to assume in principle a broad range of helical conformations, we have shown that the DNA chain generally reduces the number of base pairs per helix turn under the effect of a mechanical stretching. While the molecule over-twists in order to minimize its free energy, this behavior appears physically correlated to the shrinking of the helix diameter which is in fact consistently obtained by our method.  

These results agree with the experimental behavior,  observed for kilo-base long chains, in the low to intermediate force regime. Instead, for applied forces larger than $30$ pN, experiments show that long chains eventually untwist. To check whether the transition between over-twisting and untwisting regimes could be predicted by our model, we have considered that the amplitude of the intra-strand bending fluctuations may depend on the external force. Assuming a physically plausible force dependent cutoff on the bending fluctuation integral,  we have indeed found that also short sequences may first over-twist and then untwist by enhancing the external load. Accordingly, when the molecule untwists, its average diameter expands and the average intra-strand base pair distance grows. Although, in such short chains, the transition may occur at weaker forces than those measured in kilo-base DNA molecules, these findings suggest an intrinsic correlation between bending and twisting degrees of freedom at any length scale. However, at this stage, we do not have specific experimental information to establish the possible, quantitative dependence of the bending amplitude on the applied forces.

We have here chosen a set of model parameters appropriate to homogeneous chains of GC-base pairs as discussed in previous studies.
While the values presented in this paper for the average helical repeat and helix diameter versus applied force have some dependence on the specific input parameters, nevertheless the overall trend of our conclusions, summarized in Fig.~\ref{fig:5} for a very short sequence, is not affected by the specific parameter choice. 
On the base of the presented results we thus believe that our computational method may offer a robust tool to investigate a large variety of molecule conformations subject to external perturbations and also model the DNA response upon protein binding.


\begin{thebibliography}{widest-label}


\bibitem{travers} 
Travers A A and  Thompson J M T 2004  \textit{Phil. Trans. R. Soc. Lond. A},   \textbf{362}, 1265-1279

\bibitem{kalos11} 
Apostolaki A and  Kalosakas G  2011 \textit{Phys. Biol.},  \textbf{8}, 026006 

\bibitem{cherstvy}
Cherstvy A G  2011 \textit{Phys. Chem. Chem. Phys.},   \textbf{ 13}, 9942-9968

\bibitem{marko15}
Marko J F  2015  \textit{Physica A},  \textbf{418}, 126-153


\bibitem{radding} 
Radding C M  1991  \textit{J. Biol. Chem.},  \textbf{266}, 5355-5358

\bibitem{cox}
Cox M M  1991  \textit{Mol. Microbiol.},  \textbf{5}, 1295-1299


\bibitem{dekker}
De Vlaminck I,   van Loenhout M T J,   Zweifel L,   den Blanken J,   Hooning K,   Hage S, 
Kerssemakers J  and Dekker C 2012  \textit{Mol. Cell},  \textbf{46}, 616-624 


\bibitem{forget}
Forget A L and  Kowalczykowski S C 2012  \textit{Nature},  \textbf{482}, 423

\bibitem{stasiak}
Kiianitsa K and  Stasiak A 1997  \emph{Proc. Natl. Acad. Sci. USA},  \textbf{94}, 7837-7840


\bibitem{olson98} 
Olson W K,  Gorin A A,  Lu X -J,  Hock L M and  Zhurkin V B  1998  \emph{Proc. Natl. Acad. Sci. USA},  \textbf{95}, 11163-11168

\bibitem{marko16}
Keenholtz R A,  Grindley N D F,  Hatfull G F and  Marko J F 2016  \emph{Nucleic Acids Res.},  \textbf{44},  8921-8932 

\bibitem{chu}
Chu S 1991 \emph{Science},    \textbf{253},  861-866

\bibitem{busta92}
Smith S, Finzi L and Bustamante  C  1992  \emph{Science},   \textbf{258},  1122-1126


\bibitem{cluzel}
Cluzel P, Lebrun A,  Heller C,  Lavery R,  Viovy J L,  Chatenay D and  Caron F 1996  \emph{Science},   \textbf{271},  792-794

\bibitem{block97} 
Wang M D,  Yin H,  Landick R,  Gelles J and  Block S M  1997  \textit{Biophys. J.},    \textbf{72}, 1335-1346

\bibitem{das14}
Chou F-C, Lipfert J, Das R 2014 \textit{PLoS Comput. Biol.} \textbf{10},  e1003756 


\bibitem{busta94}
Bustamante C,  Marko J F,  Siggia E D and  Smith S  1994 \emph{Science},   \textbf{265},  1599-1601

\bibitem{odi}
Odijk T 1995  \emph{Macromolecules},   {\bf 28}, 7016-7018

\bibitem{busta96}
Smith S B,  Cui Y and  Bustamante C  1996  \textit{Science},    \textbf{271}, 795-798

\bibitem{marko97}
Marko J F   1997  \emph{Europhys. Lett.},   {\bf 38}, 183-188

\bibitem{strick98}
Strick T R, Allemand J F, Bensimon D and  Croquette  V  1998  \emph{Biophys. J.},   \textbf{74},  2016-2028

\bibitem{nelson03}
Storm C and Nelson P C 2003  \emph{Phys. Rev. E},   \textbf{67}, 051906

\bibitem{rouzina01}
Rouzina I and Bloomfield V A  2001  \emph{Biophys. J.},  \textbf{80}, 882-893; ibid., \textbf{80}, 894-900

\bibitem{cassuto} 
Cassuto E and  Howard-Flanders P 1986  \textit{Nucleic Acids Res.},  \textbf{14}, 1149-1157


\bibitem{marko99} 
L\'{e}ger J F,   Romano G,   Sarkar A,   Robert J,   Bourdieu L,   Chatenay D  and  Marko J F  1999  \emph{Phys. Rev. Lett.},  \textbf{83}, 1066-1069


\bibitem{mameren} 
van Mameren J,  Gross P, Farge  G,  Hooijman P,  Modesti M,  Falkenberg M,
Wuite G J L and  Peterman E J G  2009  \emph{Proc. Natl. Acad. Sci. USA},    \textbf{106},  18231-18236


\bibitem{busta06} 
Gore J,  Bryant Z,  N\"{o}llmann M,  Le M U,  Cozzarelli N R and  Bustamante C 2006  \textit{ Nature},  \textbf{442}, 836-839


\bibitem{croq06}
Lionnet T,   Joubaud S,   Lavery R,   Bensimon D and Croquette V  2006  \emph{Phys. Rev. Lett.},   \textbf{96}, 178102

\bibitem{mad}
Duri\v{c}kovi\`{c} B, Goriely A, Maddocks J H 2013 \emph{Phys. Rev. Lett.}, \textbf{111}, 108103

\bibitem{wuite11} 
Gross P,  Laurens N,  Oddershede L B,  Bockelmann U,  Peterman E J G and  Wuite G J L  2011  \textit{ Nature Phys.}, \textbf{7},
731–736


\bibitem{archer} 
Yuan C,   Chen H,  Lou X W and  Archer L A 2008 \textit{Phys. Rev. Lett.},    {\bf 100}, 018102

\bibitem{gole} 
Noy A and  Golestanian R 2012  \textit{Phys. Rev. Lett.},    {\bf 109}, 228101

\bibitem{tan15} 
Wu Y Y,  Bao L,  Zhang X and  Tan Z J 2015  \emph{J. Chem. Phys.},   \textbf{142}, 125103


\bibitem{fenn}
Mathew-Fenn R S,  Das R and  Harbury P A B  2008  \textit{Science},   \textbf{322}, 446-449

\bibitem{mastro} 
Mastroianni A J,  Sivak D A,  Geissler P L and  Alivisatos A P  2009  \emph{Biophys. J.},   \textbf{97},  1408-1417

\bibitem{mazur}
Mazur A K and  Maaloum M  2014  \emph{Phys. Rev. Lett.},   \textbf{112},  068104

\bibitem{kimkim16}
Kim Y -J and  Kim D -N 2016  \textit{PLoS ONE},  \textbf{11}, e0153228 


\bibitem{io09}
Zoli M   2009  \emph{Phys.Rev. E},   \textbf{79},  041927

\bibitem{io10}
Zoli M   2010  \emph{Phys.Rev. E},    \textbf{81},  051910

\bibitem{io11}
Zoli M    2011  \emph{J. Chem. Phys.},   \textbf{135},  115101

\bibitem{io16b}
Zoli M 2016 \textit{J. Chem. Phys.},   {\bf 144},  214104 

\bibitem{io16a}
Zoli M 2016 \textit{Phys. Chem. Chem. Phys.},  {\bf 18}, 17666 

\bibitem{horo}
Horowitz D S, Wang J C 1984 \textit{J. Mol. Biol.}, \textbf{173}, 75-91

\bibitem{fogg}
Fogg J M, Kolmakova N, Rees I, Magonov S, Hansma H,
Perona J J, and Zechiedrich E L 2006 \textit{J. Phys.: Condens. Matter},  \textbf{18}, S145–S159

\bibitem{irob}
Irobalieva R N,  Fogg J M,  Catanese D J,  Sutthibutpong T,
Chen M,  Barker A K,  Ludtke S J,  Harris S A,  Schmid M F,
Chiu W and Zechiedrich  L  2015  \textit{Nat. Commun.}  \textbf{6}, 8440 


\bibitem{bates}
Bates A D and  Maxwell A  2009 \emph{DNA Topology} (Oxford University Press, Oxford)

\bibitem{orozco15}
Rossetti G,  Dans P D, Gomez-Pinto I,  Ivani I,  Gonzalez C and  Orozco M   2015 \textit{Nucleic Acid Res.}, \textbf{43},  4309–4321

\bibitem{wang15}
Li G,  Shen H, Zhang D, Li Y, and Wang H 2016 \textit{J. Chem. Theory Comput.}, \textbf{12},  676-693

\bibitem{nina10}
Szatylowicz H and Sadlej-Sosnowska  N 2010  \textit{J. Chem. Inf. Model.},  \textbf{50}, 2151–2161

\bibitem{gueron}
Gu\'{e}ron M, Kochoyan M and Leroy J L  1987  \textit{Nature},   \textbf{ 328}, 89-92

\bibitem{benham}
Fye R M and  Benham C J    1999   \emph{Phys. Rev. E},  {\bf 59}, 3408-3426


\bibitem{bonnet}
Altan-Bonnet G, Libchaber A and  Krichevsky O 2003  \textit{Phys.\ Rev.\ Lett.},   {\bf 90}, 138101

\bibitem{russu}
Chen C and  Russu I M  2004  \emph{Biophys. J.},  \textbf{87}, 2545-2551

\bibitem{rapti}
Rapti Z,  Smerzi A,  Rasmussen K {\O}, Bishop A R, Choi  C H and  Usheva A  2006  \emph{Phys. Rev. E},  \textbf{73},   051902

\bibitem{metz09}
Metzler R, Ambj\"{o}rnsson T, Hanke A and Fogedby H C 2009  \emph{J. Phys.: Condens. Matter} \textbf{21} 034111

\bibitem{bishop09}
Alexandrov B,  Voulgarakis N K,  Rasmussen K {\O},
Usheva A and  Bishop A R  2009 \textit{J. Phys.: Condens. Matter}  {\bf 21},  034107


\bibitem{kame06}
Krueger A,  Protozanova E and Frank-Kamenetskii M D  2006 \emph{Biophys. J.},   \textbf{90},  3091-3099


\bibitem{porsch} 
Porschke D  1991  \textit{Biophys. Chem.},  \textbf{40}, 169

\bibitem{menon}
Padinhateeri R and Menon G I  2013  \emph{Biophys. J.},   \textbf{104},  463-471

\bibitem{manghi15} 
Brunet A, Tardin C, Salom\'{e} L,  Rousseau P,  Destainville N and  Manghi M 2015  \emph{Macromolecules},  \textbf{48}, 3641-3652


\bibitem{ejte15}
Salari H, Eslami-Mossallam B, Naderi S and  Ejtehadi M R  2015  \emph{J. Chem. Phys.},    \textbf{143},  104904


\bibitem{kim16} 
Waters J T and  Kim H D 2016  \textit{Phys. Rev. E},    \textbf{93}, 043315


\bibitem{proho}
Gao Y, Devi-Prasad K V and  Prohofski E W 1984  \emph{J. Chem. Phys.},    \textbf{80}, 6291

\bibitem{joy09}
Joyeux M and Florescu A -M 2009 \textit{J. Phys.: Condens. Matter}  {\bf 21},  034101

\bibitem{weber09}
Weber G, Haslam N, Essex J W and  Neylon C   2009 \textit{J. Phys.: Condens. Matter} \textbf{21}, 034106

\bibitem{baird}
Baird N C  1974  \textit{Int. J. Quantum Chem. },  \textbf{8}, (S1) 49-54

\bibitem{pey2}
Dauxois T, Peyrard M and Bishop  A R 1993  \emph{Phys. Rev. E},   \textbf{47},  R44-47


\bibitem{io12}
Zoli M  2012  \textit{J. Phys.: Condens. Matter},    {\bf 24},  195103


\bibitem{sims15}
Lambropoulos K, Chatzieleftheriou M,   Morphis A,  Kaklamanis K, Theodorakou M,
and  Simserides C 2015 \textit{Phys. Rev. E}, \textbf{92}, 032725


\bibitem{sims16}
Lambropoulos K, Chatzieleftheriou M,   Morphis A,  Kaklamanis K,  Lopp R, Theodorakou M,
Tassi M and  Simserides C 2016 \textit{Phys. Rev. E}, \textbf{94}, 062403

\bibitem{io14b} 
Zoli M  2014 \textit{J. Theor. Biol.},   {\bf 354},  95-104 


\bibitem{fehi}
Feynman R P and  Hibbs  A R 1965 {\it Quantum Mechanics and Path Integrals}, (Mc Graw-Hill, New York)

\bibitem{io11a}
Zoli M   2011  \textit{Eur. Phys. J. E},   {\bf 34}, 68

\bibitem{io03}
Zoli M 2003  \textit{Phys. Rev. B},  {\bf 67}, 195102

\bibitem{io13}
Zoli M 2013  \textit{J. Chem. Phys. },  {\bf 138}, 205103


\bibitem{crick}
Crick F H and  Klug A  1975 \emph{Nature},  \textbf{255}, 530-533

\bibitem{zocchi13}
Sanchez D S, Qu H,  Bulla D and  Zocchi G 2013  \textit{Phys. Rev. E},  \textbf{87}, 022710

\bibitem{harris} 
Mitchell J S, Laughton C A and  Harris  S A 2011  \textit{Nucleic Acids Res.},  \textbf{39}, 3928-3938


\bibitem{kim14}
Le T T and  Kim H D 2014  \emph{Nucl. Acids Res.},    \textbf{42}, 10786-10794

\bibitem{soder} 
Ullner M,  J\"{o}nsson B,  Peterson C,  Sommelius O and  S\"{o}derberg  B 1997  \emph{J. Chem. Phys.},   \textbf{107}, 1279

\bibitem{volo10}
Geggier S and  Vologodskii A 2010  \emph{Proc. Natl. Acad. Sci. USA},   \textbf{107}, 15421-15426

\bibitem{weber15}
Ferreira I, Amarante T D and Weber  G 2015 \emph{J. Chem. Phys.},   \textbf{143}, 175101

\bibitem{wang}
Wang J C 1976 \emph{Proc. Natl. Acad. Sci. USA},   \textbf{76},  200-203

\bibitem{levitt78}
Levitt M  1978 \emph{Proc. Natl. Acad. Sci. USA},  \textbf{75}, 640-644

\bibitem{hayes}
Hayes J J, Tullius T D and  Wolffe A P 1990 \emph{Proc. Natl. Acad. Sci. USA},  \textbf{87}, 7405-7409

\bibitem{duguet}
Duguet M   1993 \emph{Nucleic Acids Res.}, \textbf{21}, 463-468

\bibitem{note}
For the first base pair in the chain, $< r_1 >$ is calculated with the Boltzmann weight factor associated to the action $A_a[r_1]$ in Eq.~(\ref{eq:02}) which does not depend on angular variables.

\bibitem{peters13} 
Peters J P,  Yelgaonkar S P,  Srivatsan S G,  Tor Y and  Maher L J   2013 \textit{Nucleic Acid Res.}, \textbf{41}, 10593-10604

\bibitem{olson99}
Kosikov K M,  Gorin A A,  Zhurkin V B and  Olson W K 1999  \textit{J. Mol. Biol.},  \textbf{289}, 1301-1326

\bibitem{zachar15}
Liebl K,  Drsata T,  Lankas F,  Lipfert J and  Zacharias  M  2015 \textit{Nucleic Acid Res.},  \textbf{43},  10143-10156









\end{thebibliography}
\end{document}